\documentclass[aps,prb,a4paper,twocolumn,superscriptaddress,nobibnotes]{revtex4-1}
\usepackage{helvet}
\usepackage{graphicx}
\usepackage{xspace}
\usepackage{amsmath,amssymb,bm}
\usepackage[usenames,dvipsnames]{xcolor}

\newif\ifdrafttext
\drafttextfalse 
\ifdrafttext \usepackage[colorlinks,urlcolor=black,citecolor=black,linkcolor=black]{hyperref} \else   \usepackage[hidelinks]{hyperref} \fi

\newcommand{\titleofpaper}{Flux-tunable heat sink for quantum electric circuits}

\newcommand{\QCDaffiliation}{QCD Labs, COMP Centre of Excellence, Department of Applied Physics, Aalto University, P.O. Box 13500, FI-00076 Aalto, Finland}
\newcommand{\NISTaffiliation}{National Institute of Standards and Technology, Boulder, Colorado, 80305, USA}
\newcommand{\VTTaffiliation}{VTT Technical Research Centre of Finland, P.O. Box 1000, FI-02044 VTT, Finland}
\newcommand{\MSPaffiliation}{MSP group, COMP Centre of Excellence, Department of Applied Physics, Aalto University, P.O. Box 13500, FI-00076 Aalto, Finland}
\newcommand{\OULUaffiliation}{Nano and Molecular Systems Research Unit, University of Oulu, P.O. Box 3000, FI-90014 Oulu, Finland}
\newcommand{\LOUGHBOROUGHaffiliation}{Departments of Mathematical Sciences and Physics, Loughborough University, Loughborough, Leicestershire LE11 3TU, United Kingdom}
\newcommand{\BROWNaffiliation}{Department of Physics, Brown University, Box 1843, Providence, Rhode Island 02912-1843, USA}

\begin{document}

\title{\titleofpaper}

\author{M. Partanen}
\affiliation{\QCDaffiliation}
\author{ K. Y. Tan}
\affiliation{\QCDaffiliation}
\author{ S. Masuda}
\affiliation{\QCDaffiliation}
\author{ J. Govenius}
\affiliation{\QCDaffiliation}
\author{ R. E. Lake}
\affiliation{\QCDaffiliation}
\affiliation{\NISTaffiliation}
\author{ M. Jenei}
\affiliation{\QCDaffiliation}
\author{ L. Gr\"onberg}
\affiliation{\VTTaffiliation}
\author{ J. Hassel}
\affiliation{\VTTaffiliation}
\author{ S. Simbierowicz}
\affiliation{\VTTaffiliation}
\author{ V. Vesterinen}
\affiliation{\QCDaffiliation}
\affiliation{\VTTaffiliation}
\author{ J. Tuorila}
\affiliation{\QCDaffiliation}
\affiliation{\MSPaffiliation}
\affiliation{\OULUaffiliation}
\author{ T. Ala-Nissila}
\affiliation{\MSPaffiliation}
\affiliation{\LOUGHBOROUGHaffiliation}
\affiliation{\BROWNaffiliation}
\author{ M. M\"ott\"onen}
\affiliation{\QCDaffiliation}

\date{\today}

\begin{abstract}

Superconducting microwave circuits show great potential for practical quantum technological applications such as quantum information processing.
However, fast and on-demand initialization of the quantum degrees of freedom in these devices remains a challenge.
Here, we experimentally implement a tunable heat sink that is potentially suitable for the initialization of superconducting qubits.
Our device consists of two coupled resonators.
The first resonator has a high quality factor and a fixed frequency  whereas the second resonator is designed to have  a low quality factor and a tunable resonance frequency.
We engineer the low quality factor using an on-chip resistor and the frequency tunability using a superconducting quantum interference device.
When the two resonators are in resonance, the photons in the high-quality resonator can be efficiently dissipated.
We show that the corresponding loaded quality factor can be tuned from above $10^5$ down to a few thousand at 10~GHz in  good  quantitative agreement with our theoretical model.

\end{abstract}

\maketitle
\newpage

\renewcommand{\figurename}{Figure}
\renewcommand{\tablename}{Table}

\section*{Introduction}

One of the most promising approaches to building a quantum computer is based on superconducting qubits in the framework of circuit quantum electrodynamics~\cite{Ladd, Clarke, Blais, Wallraff_Nature_2004, Goppl,Kelly2015}. 
However, not all of the criteria  for a functional quantum computer~\cite{DiVincenzo2000} have  been achieved simultaneously  at the desired level. 
In particular,  computational errors need to be mitigated with quantum error correction~\cite{Lidar_book_2013,Terhal_2015}.
Many quantum error correction codes require frequent initialization of ancillary qubits during the computation.
Thus, fast and accurate qubit reset is a typical requirement in the efficient implementation of quantum  algorithms.
To date, several approaches for qubit initialization have been studied~\cite{Valenzuela_2006, Johnson_2012, Riste_2012, Geerlings_2013, Bultink_2016}. 
Initialization to the ground state by waiting is a straightforward method but it becomes impractical in repeated measurements of qubits with long lifetimes. 
Therefore, active initialization is advantageous.
Furthermore, it may be beneficial to design individual circuits for qubit control, readout, and initialization in order to avoid performance-limiting compromises in the optimization of the circuit parameters. 
In this work we focus on a specialized initialization circuit, which remains to be implemented in superconducting quantum processors.

Recently, a promising qubit initialization protocol based on dissipative environments was proposed in Refs.~\onlinecite{Jones_3, Tuorila_2017}.
In this proposal,  a resistor coupled to a frequency-tunable resonator quickly absorbs the excitation from the qubit when tuned in resonance. 
In this paper, we experimentally realize such a tunable dissipative environment and study its effect on a superconducting resonator.
Tunable superconducting resonators  have been demonstrated previously~\cite{Palacios-Laloy2008, Healey_2008, Pierre_2014, Wang_2013, Vissers_2015, Adamyan_2016} but without engineered dissipation arising from on-chip normal-metal components. 
In addition to quantum computing, very sensitive cryogenic detectors~\cite{Inomata_2016,Govenius_2016,Narla_2016} may benefit from tunable dissipation for calibration purposes.
Furthermore, tunable transmission lines are also useful in studying fundamental  quantum phenomena~\cite{Wilson_2011}.

Although dissipation is in some cases beneficial for quantum computing~\cite{Verstraete2009}, lossy materials are typically  harmful for qubit lifetimes during computation. 
Therefore, one needs to be able to switch the dissipation on and off deterministically.
In state-of-the-art experiments, quality factors, $Q$, above $10^6$ indicating very low dissipation have been achieved with coplanar-waveguide  resonators~\cite{Megrant_2012}. 
Various materials and  methods have been studied for fabricating high-$Q$ resonators~\cite{ Vissers_2010, Sandberg_2012, Bruno_2015, Zmuidzinas_2012}.
Here we  fabricate  high-$Q$ resonators based on niobium  on a silicon wafer.
In addition, we tune the  $Q$ factor from above $10^5$ down to a few thousand by coupling the resonator relatively strongly to a dissipative element.
Importantly, the integrated resistive element we introduce does not inherently degrade the $Q$ factor when it is  weakly coupled to the resonator compared with similarly fabricated resonators without any engineered resistive elements.

\begin{figure*}[t]
\centering
\includegraphics[width=\linewidth]{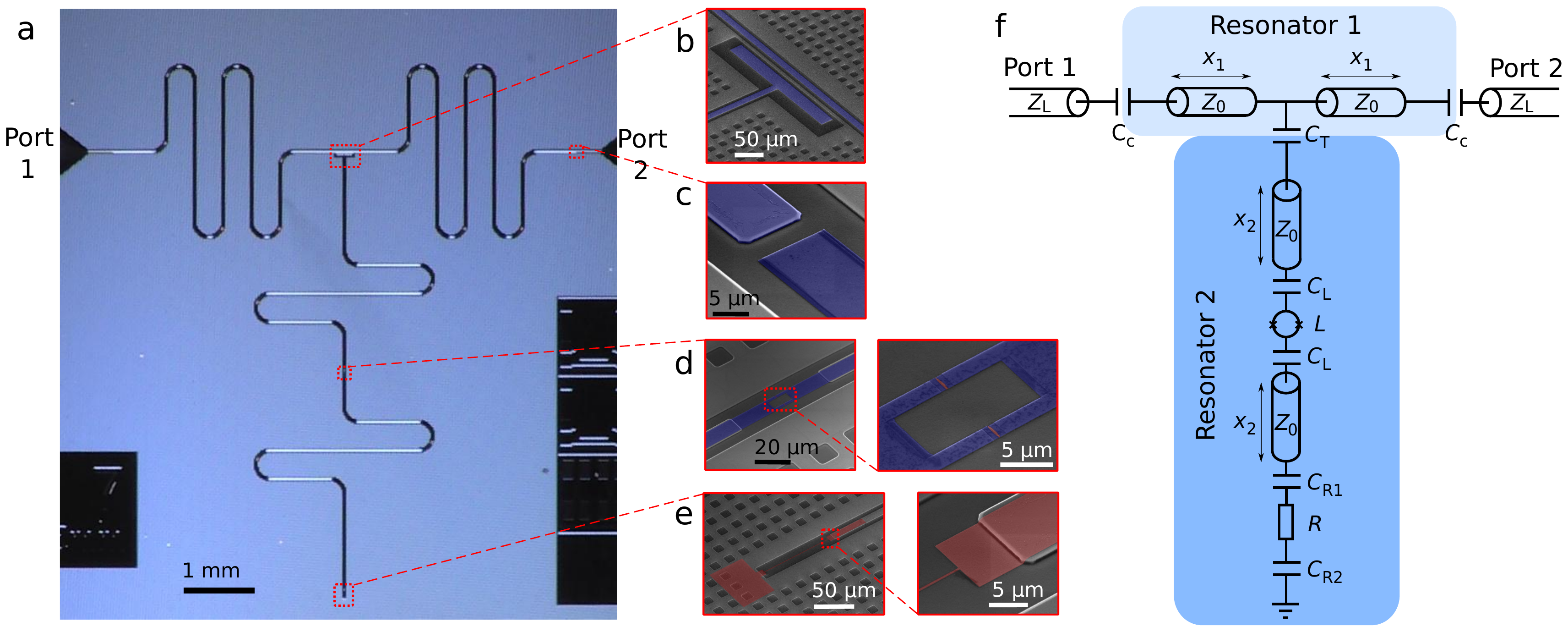}
\caption{Sample structure. 
(a) Optical top image of the measured sample. 
(b) False-colour scanning electron microscope  image of the coupling capacitor between the two resonators, 
and (c) between Resonator 1 (light blue) and the port to the external transmission line (dark blue).
(d) Two micrographs of the SQUID loop highlighted in blue and the junctions highlighted in red.
(e) Two micrographs of the termination Cu resistor (red).
(f) Electrical circuit diagram of the sample.
Resonators 1 and 2 with characteristic impedances $Z_0$ are coupled to each other by a coupling capacitance $C_\textrm{T}$ and to external transmission lines with characteristic impedance $Z_\textrm{L}$  by capacitances $C_\textrm{C}$.
The inductance of the SQUID is denoted by $L$, and the termination resistance by $R$.
The SQUID is connected to the centre conductor of Resonator~2 line with capacitances $C_\textrm{L}$, and the resistor to centre conductor and ground  with  $C_\textrm{R1}$ and  $C_\textrm{R2}$, respectively.
The lengths of the resonator sections are denoted by $x_{1/2}$.
The image in panel (a) is from Sample~A, and those in panels (b)--(e) from Sample~B.
}
\label{fig:sample_structure}
\end{figure*}

\begin{table*}[t]
 \begin{center}
 \caption{ Simulation parameters. 
See Fig.~\ref{fig:sample_structure}f and text for the definition of the symbols.
Samples A and B  have the same parameter values except for the length $x_2$, where the value for Sample B is given in parenthesis.
 }
 \begin{tabular}{ |c || c | c | c | c | c | c | c | c | c | c | c | c | c | c | }
\hline
   Parameter 
& $C_\textrm{C}$ 
& $C_\textrm{T}$ 
& $C_\textrm{L}$ 
& $C_\textrm{R1}$ 
& $C_\textrm{R2}$ 
& $C_\textrm{l}$ 
& $R$   
& $Z_0$   
& $Z_\textrm{L}$   
& $\varepsilon_\textrm{eff}$ 
& $x_1$ 
& $x_2$ 
& $Q_\textrm{int,1}$ 
& $I_0$ 
\\
& (fF)
& (fF)  
& (pF) 
& (pF)
& (pF)
& (pF/m) 
& ($\Omega$)
& ($\Omega$)
& ($\Omega$) 
&  
& (mm) 
& (mm)
&  
& (nA) 
\\
\hline
\hline 
& & & & & & & & & & &  & & & \\[-1em]
Value
& 1  
& 5  
& 2.8 
& 4.0 
& 28 
& 180 
& 375 
& 50 
& 50  
& 6.35 
& 12  
& 7.5 (8.0) 
& $1\times10^{5}$ 
& 255  
\\
\hline
\end{tabular}
 \label{tab:simulation_parameters}
 \end{center}
\end{table*}

\section*{Results}

\noindent
\textbf{Experimental samples} \\
\noindent
The  structure of our device is presented in Fig.~\ref{fig:sample_structure} together with the corresponding electrical circuit diagram which defines the symbols used below.
The device consists of two coupled resonators, Resonator~1 with a fundamental frequency of 2.5~GHz, and Resonator~2 with a tunable frequency. 
Both ends of Resonator~1 couple capacitively ($C_\textrm{C}$) to external circuitry for scattering parameter measurements. 
The even harmonics of Resonator~1  interact  with Resonator~2 since there is a voltage antinode   at the center of the half-wave Resonator~1, and hence, the capacitive ($C_\textrm{T}$) coupling to Resonator~2 is significant.

The resonators are fabricated out of niobium in a coplanar-waveguide   geometry.
The modes of Resonator~2 are tunable owing to a superconducting quantum interference device (SQUID) acting as a flux-tunable inductance, placed in the middle of the resonator.
The SQUID is integrated into the center pin of the waveguide and consists of two aluminium layers separated by an  insulating aluminium oxide layer.
When the resonance frequencies of the two resonators meet, we expect a degradation of the Resonator~1 quality factor because Resonator~2 is terminated with a dissipative on-chip resistor made of copper. 
Importantly, the device is designed to retain a high quality factor of Resonator~1 whenever Resonator~2 is far detuned.

We study two samples, Sample A and B, which are  nominally identical, except for the length of Resonator~2.
We mostly focus on Sample~A which has a wider tuning range of the  quality factor of Resonator 1.
The samples are measured at a cryostat temperature of approximately 10~mK.
The theoretical model described in Methods reveals all the essential features of the two samples with a single set of parameters given in Table~\ref{tab:simulation_parameters}. 
See Methods for the details of the sample fabrication.

\begin{figure*}[t]
\centering
\includegraphics[width=\linewidth]{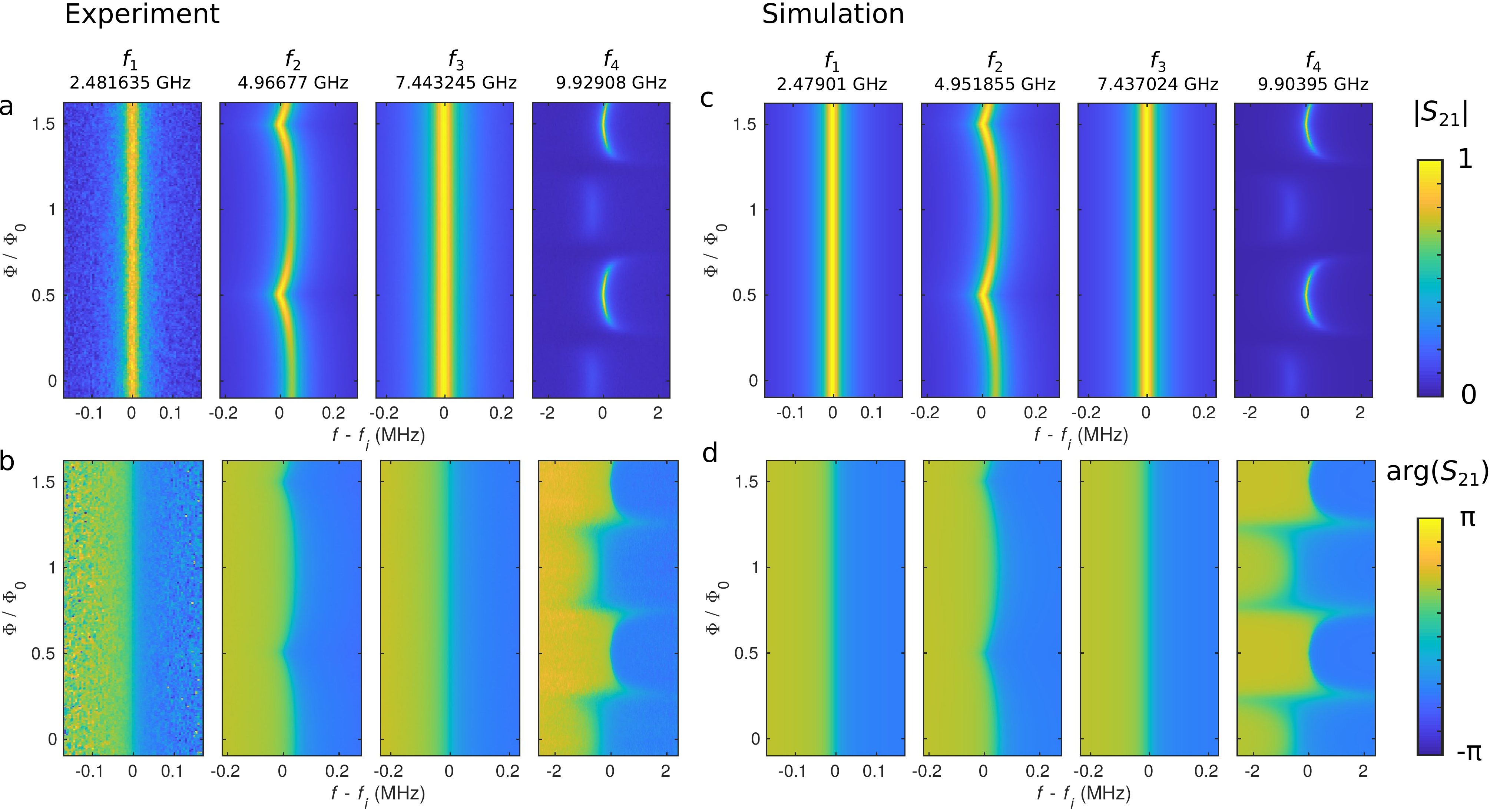}
\caption{Resonances of Sample~A. 
(a,b) Experimental and (c,d) computational (a,c) amplitude and (b,d) phase of the scattering parameter $S_{21}$ for the first four  modes  of Resonator 1  as  functions of frequency and magnetic flux.
The amplitude of $S_{21}$ in each subpanel is normalized independently by dividing with the corresponding maximum amplitude.
The power in the measurements is approximately $-90$~dBm at Port~1.
The resonance frequencies at half flux quantum are given above the panels, and the simulation parameters are given in Table~\ref{tab:simulation_parameters}.
}
\label{fig:meas_sim_modes_i7}
\end{figure*}

\begin{figure*}[t]
\centering
\includegraphics[width=0.9\linewidth]{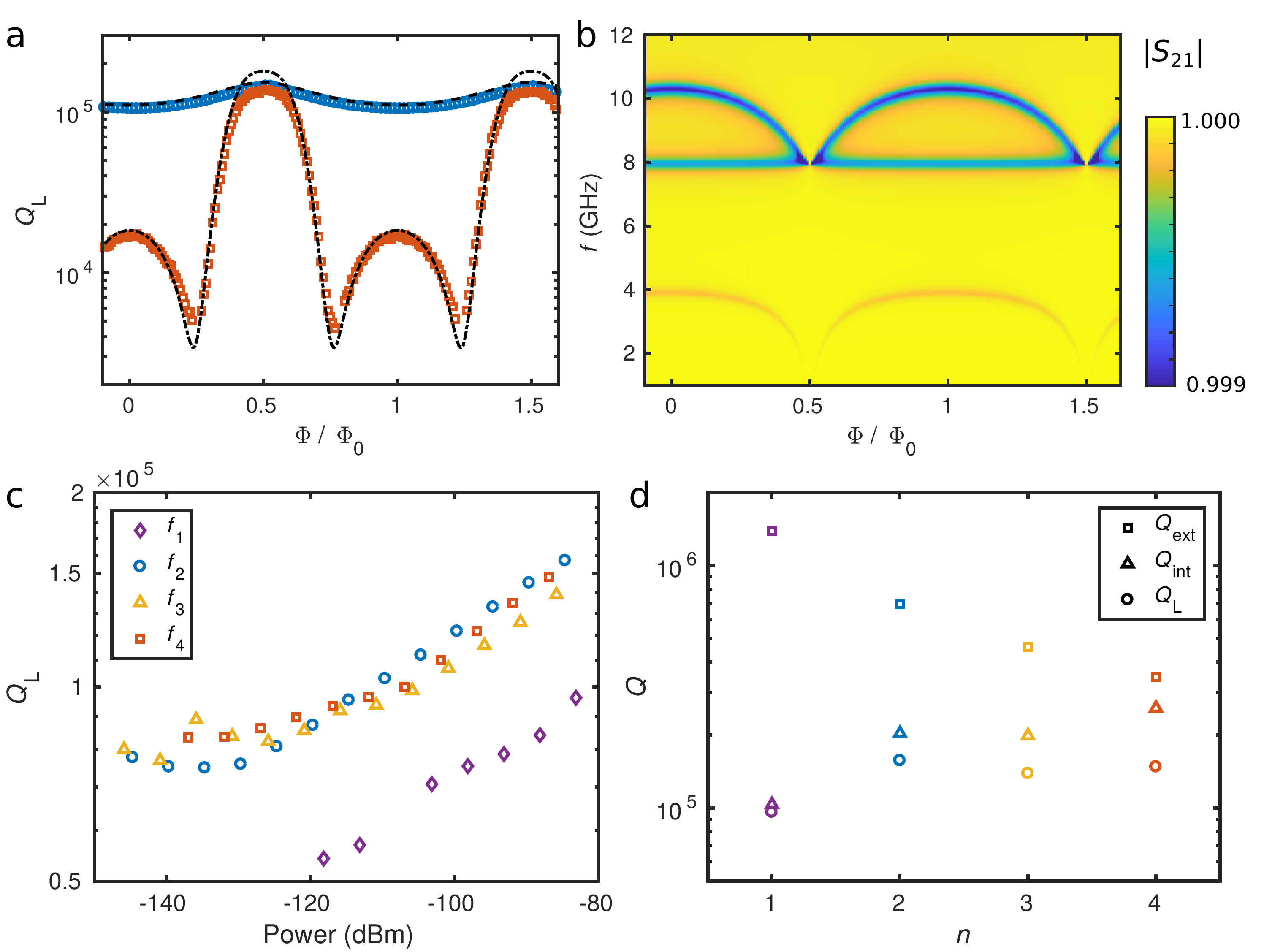}
\caption{Quality factors of Resonator~1 and resonances of Resonator~2 for Sample~A.
(a) Measured loaded quality factor, $Q_\textrm{L}$, for mode 2 (blue circles) and for mode 4 (red squares) as functions of the magnetic flux through the SQUID together with the simulated values (dashed line and dash-dotted line, respectively). 
(b) Absolute value of the simulated scattering parameter $S_\textrm{21}$ of Sample~A with only Resonator~2, i.e., at the limit $C_\textrm{C}\rightarrow \infty$. 
The colour bar is truncated at 0.999 for clarity.
(c) Measured loaded quality factor, $Q_\textrm{L}$, of Sample~A  (markers) for the first four modes as functions of power at Port~1. 
(d) Measured $Q_\textrm{L}$ of Sample~A  (circles), predicted  external quality factor, $Q_\textrm{ext}$, (squares) and calculated internal quality factor,   $Q_\textrm{int}$, (triangles) as functions of the mode number. 
The simulation parameters are given in Table~\ref{tab:simulation_parameters}.
In (a), the power at Port~1 is approximately $-90$~dBm, and in (d)  $-85$~dBm.
In (c) and (d),  the magnetic flux through the SQUID is $\Phi/\Phi_0=0.5$. 
}
\label{fig:Qfactor_i7}
\end{figure*}

\begin{figure*}[t]
\centering
\includegraphics[width=0.85\linewidth]{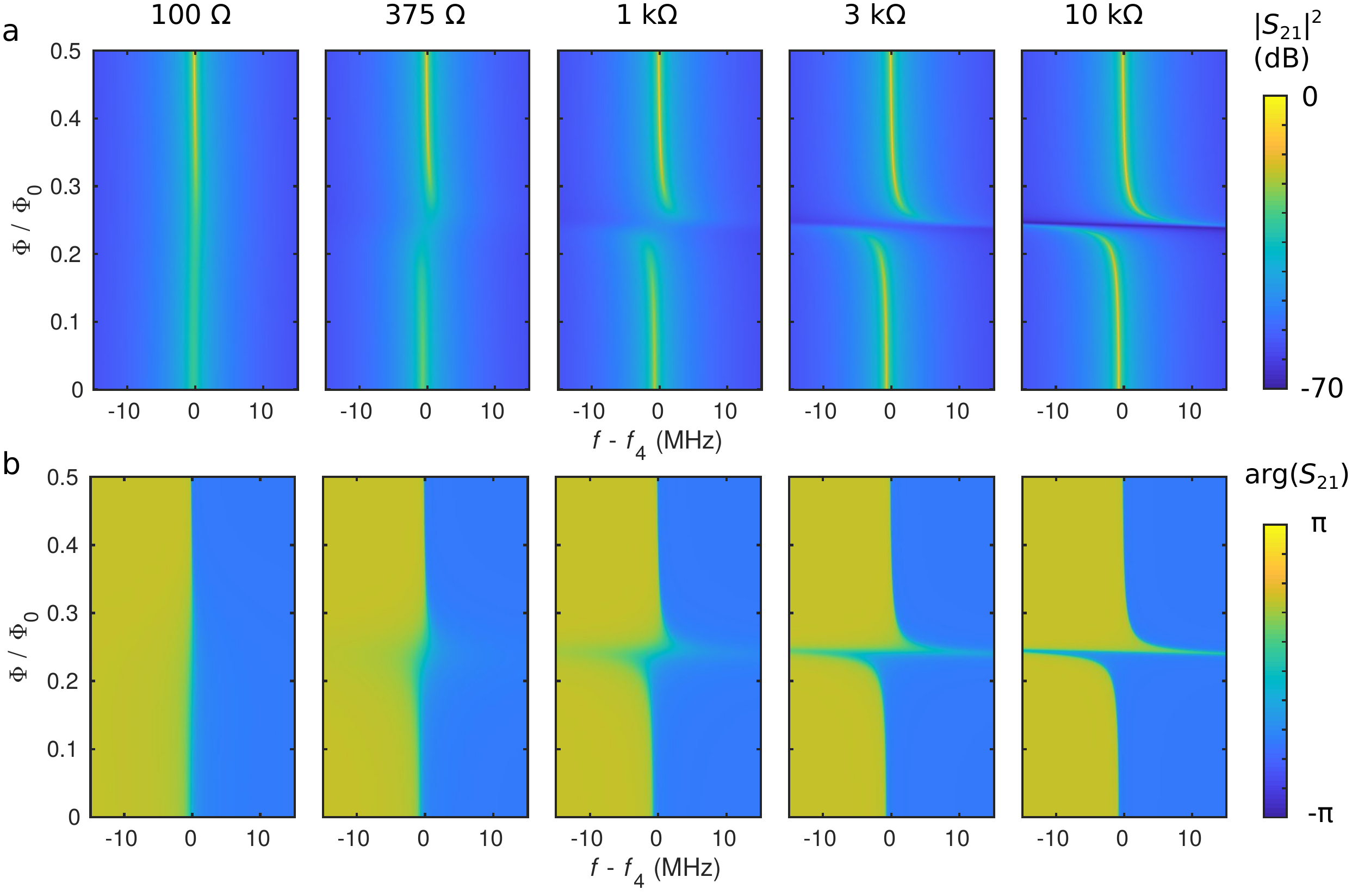}
\caption{Effect of the  termination resistance. Simulated (a) amplitude, and (b) phase of the mode 4 in Sample~A as functions of frequency and magnetic flux with different resistance values, $R$, as indicated above the panels.
The resonance frequency is $f_4=9.90395$~GHz, and the other parameters are given in Table~\ref{tab:simulation_parameters}.
}
\label{fig:sim_i7_differentR}
\end{figure*}

\noindent
\textbf{Flux dependence of the resonance frequencies}\\
\noindent
The first four resonances of Resonator~1 in Sample~A are shown in Fig.~\ref{fig:meas_sim_modes_i7}  as a function of the magnetic flux through the SQUID.
The first and the third mode at approximately 2.5 and 7.5~GHz, respectively,  do not depend on the flux due to a voltage node in the middle of Resonator~1, i.e., at  the coupling capacitor $C_\textrm{T}$.
Thus, these modes are decoupled from those of Resonator~2. 
In contrast, the second and the fourth mode at 5 and 10~GHz, respectively,  show clear flux dependence owing to the changing SQUID inductance, which in turn changes the frequencies of the modes in Resonator~2. 
If a dissipative mode in Resonator~2 approaches the frequency of a mode in Resonator~1, we observe two distinctive features: the resonance in Resonator~1  shifts and broadens owing to the coupling to the dissipative mode.
The experimental scattering parameter $S_{21}$ is normalized as explained in Methods.
The simulation based on the theoretical model (see Methods)  shows excellent agreement with experimental data. 
The slight discrepancy between the experiment and the simulation mainly arises from the uncertainty in the exact values of the parameters given in Table~\ref{tab:simulation_parameters}.

The distinctively different flux dependence of  modes 2 and 4 in Sample~A is clarified by  Fig.~\ref{fig:Qfactor_i7}b, which shows simulated  $|S_{21}|$ with only Resonator 2, i.e., in the limit $C_\textrm{C}\rightarrow \infty$. 
Resonator~2 has a flux dependent resonance near 4~GHz, which does not cross the second mode of Resonator~1 at 5~GHz.
Nevertheless, it comes sufficiently near 5~GHz, which explains the frequency shift of  mode 2 of Resonator~1.
In contrast, Resonator~2 has a flux dependent resonance near 10~GHz, very close to mode 4 of Resonator~1.
The resonances intersect which results in dramatic changes in the fourth mode of Resonator~1.
The second mode of Resonator~2 near 8~GHz has a current node at the center of the resonator, where the SQUID is located; thus, it is only very weakly dependent on the flux.

Supplementary Fig.~\ref{fig:meas_sim_modes_i8} shows results similar to those in Fig.~\ref{fig:meas_sim_modes_i7} for Sample A but for  modes 2, 3, and 4 of Sample~B. 
The simulations and experiments are also here in  good agreement. 
However, the simulated mode~2 is substantially narrower than the experimental one.
This broadening may arise from an unaccounted mode of the sample holder at a nearby frequency.
Furthermore, there is some discrepancy in the phase of  mode~4 near integer flux quanta. 
This discrepancy can be explained by  uncertainty in the normalization procedure with very small amplitudes.
The first mode is outside the frequency range of the used microwave components, and hence we do not show data for it. 
For a quantitative comparison of the measured and the simulated resonance frequencies in Samples~A and B, Supplementary Fig.~\ref{fig:meas_sim_delta_f} shows the frequency shifts of modes 2 and 4 extracted from Fig.~\ref{fig:meas_sim_modes_i7} and Supplementary Fig.~\ref{fig:meas_sim_modes_i8}.
The flux dependence of the modes of Resonator~2  is similar in Sample~B to that of Sample~A as  shown in Supplementary Fig.~\ref{fig:Qfactor_i8}b.
However, the resonances do not  intersect at 10~GHz although they are very close to each other.

\noindent
\textbf{Quality factors}\\
\noindent
We also analyze the quality factors as functions of flux, as shown for Sample~A in Fig.~\ref{fig:Qfactor_i7}a, and for Sample~B in Supplementary Fig.~\ref{fig:Qfactor_i8}a.
The $Q$ factors of the second and fourth mode are tunable unlike in the case of the first and third mode.
The second mode of Sample~A shows only relatively small variation near $10^5$ whereas the fourth mode can be tuned from above  $10^5$ down to a few thousand.
For Sample~B, the flux dependence of the $Q$ factor is similar.
However, the second mode has a substantially lower experimental loaded quality factor, $Q_\textrm{L}$, than the simulated value, i.e., a broader resonance peak as discussed above.
A better agreement between the simulation and the experiment can be obtained by introducing an additional loss mechanism as described in the caption of Supplementary Fig.~\ref{fig:Qfactor_i8}.

The power dependence of  the quality factors is analyzed in Fig.~\ref{fig:Qfactor_i7}c for the four lowest  modes in Sample~A.
The $Q$ factors decrease with decreasing power as expected~\cite{Zmuidzinas_2012}. 
Nevertheless, they remain rather close  to  $10^5$ even at the single-photon level, around -140~dBm.
However, relatively high powers enable more accurate measurements of the losses caused by the resistor when  the resonators are tuned into the weak coupling regime.
Figure~\ref{fig:Qfactor_i7}d shows the experimentally obtained loaded quality factor, $Q_\textrm{L}$, and the theoretically predicted external quality factor, $Q_\textrm{ext}$, corresponding to the losses through the coupling capacitors $C_\textrm{C}$ as functions of the mode number $n$.
Furthermore, the internal quality factor, $Q_\textrm{int}$, corresponding to the internal losses in the system is calculated  from the equation $Q_\textrm{int}^{-1} = Q_\textrm{L}^{-1} - Q_\textrm{ext}^{-1}$.
The internal quality factor slightly increases with the mode number and obtains values  near $2.5\times10^5$.
The minimum value of $Q_\textrm{L} \lesssim  5\times10^3$ in Fig.~\ref{fig:Qfactor_i7}a gives also the minimum value for $Q_\textrm{int}$ since the internal losses of the system dominate when the resistor is strongly coupled to the fourth mode of Resonator~1. 

The minimum and maximum $Q_\textrm{int}$ correspond to photon lifetimes $\tau_\textrm{int}= Q_\textrm{int}/\omega_0$ of  80~ns  and 4~$\mu$s, respectively, at $\omega_0 =2 \pi \times10$~GHz when other losses are neglected.
Furthermore,  $Q_\textrm{ext}$ corresponds to a photon lifetime of 6~$\mu$s.
These photon lifetimes are long compared to the period of the coherent oscillations  between the two resonators at resonance, $\tau_\textrm{T}=30$~ns (see Methods).
Thus, the internal or external losses of Resonator~1 are not dominating over the coupling strength between the resonators.
However, the simulated $Q$ factors of the lowest modes of Resonator~2 in  Fig.~\ref{fig:Qfactor_i7}b  are well below 40 at the zero flux bias and also at $\Phi/\Phi_0 \approx 0.2$ which corresponds to the crossing of the modes at 10~GHz.
They obtain values above 100 only in the range $0.48 <\Phi/\Phi_0< 0.52$ due to the ideally diverging SQUID inductance.
Thus, the photon lifetime  in Resonator~2 is below 0.6~ns at 10~GHz and at the relevant flux point.
Consequently, the photons in Resonator~2 are dissipated quickly compared to the period of the coherent oscillations between the resonators, which prevents the formation of well-separated modes hybridized between the resonators.
Importantly, Resonator~2 mostly functions as a tunable dissipative environment for Resonator~1, the dissipation of which is limited by the coupling strength between the resonators.

\noindent
\textbf{Simulations with different resistances}\\
\noindent
We also simulate the effect of  changing the termination resistance as shown in Fig.~\ref{fig:sim_i7_differentR}. 
The other parameters in the simulations are from Sample~A. 
Note the different frequency range and colour scale compared to Fig.~\ref{fig:meas_sim_modes_i7}.
In the case of a 100-$\Omega$ termination resistance, there is very little shift in the resonance frequency as a function of the magnetic flux.
Nevertheless, the width of the peak varies since the ideal SQUID inductance diverges, $L \rightarrow \infty$  for $\Phi/\Phi_0 \rightarrow 0.5$, and therefore, it decouples the resistor from Resonator~1.
At even lower resistances near 50~$\Omega$ (not shown), the termination is well matched to the characteristic impedance, and hence the description of Resonator~2 as a resonator becomes obscure. 
Instead, it appears as a broad-band dissipative environment for Resonator~1.
With increasing resistance, Resonator~2 obtains  well-defined resonances, with zero-flux $Q$ factors becoming of the order of $10^3$ at  $R=10$~k$\Omega$.
However, the maximum $Q_\textrm{L}$  in Resonator~1 of $1.8\times10^5$ does not vary due to the ideally infinite impedance of the SQUID at $\Phi/\Phi_0 \rightarrow 0.5$.
In contrast  at zero flux, $Q_\textrm{L}$  increases from  $1.0\times10^4$ to  $1.3\times10^5$  as the resistance  increases from  $100$~$\Omega$ to  $10$~k$\Omega$.
At $R=10$~k$\Omega$, the  two resonators show a clear avoided-crossing feature.
There is a continuous crossover from a single modulating  resonance at low resistance values  to two resonances with an avoided crossing at high resistances. 
In the experiments, we have  $R=375$ $\Omega$, which results in a single modulating  resonance with some avoided-crossing-like features.

\section*{Discussion}

We have experimentally demonstrated  tunable dissipation in a  device consisting of two resonators in  very good agreement with our theoretical model. 
We have studied two samples with  slightly different parameters.
Both of them allow us to substantially tune the loaded quality factors of the relevant resonances.
In addition, the internal quality factor of one of the modes   can be tuned from approximately a quarter of a million down to a few thousand.
Importantly, we have designed the circuit such that the coupling strength between the resonators is  somewhat weaker than the dissipation in Resonator~1 and stronger than the dissipation in Resonator~2.
Therefore, Resonator 2 operates as an efficient dissipative environment for Resonator 1.
Note that the spurious internal losses in the system are low as indicated by the  high maximum quality factor.
Thus, the fabrication of the on-chip resistors is compatible with obtaining high quality factors using our fabrication process.
To our knowledge, these are the highest demonstrated quality factors in superconducting resonators with integrated on-chip resistors. 
In the future, the remaining unwanted  losses can be reduced by further improving the process.

Here, we have demonstrated a  tunable dissipative environment with   a rather specific sample type.
Nevertheless, the geometry and parameters can be relatively freely chosen to optimize the heat sink  for different applications.
For instance, it is possible to modify the losses by changing the resistance and capacitance values.
Furthermore, the geometry of the system can be changed in order to obtain different  coupling strengths  for different modes.
In addition, the resistor does not necessarily have to be directly coupled to Resonator~2. 
Instead, it can be outside the resonator and coupled with a small capacitance and  a section of a transmission line. 
Furthermore, in case the resistance equals to the characteristic impedance of the transmission line resonator, the environment is effectively similar to a transmission line~\cite{Pierre_2014}.

Although we consider the resistors only as  sources of dissipation here, they may also be engineered to simultaneously function in photon-absorbing normal-metal--insulator--superconductor tunnel junctions~\cite{Tan_2017}, or  quasiparticle traps in superconducting circuits~\cite{Goldie_1990,Ullom_2000,Rajauria_2012,Riwar_2016,Patel_2017}.
Fast tuning of the quality factors can be obtained by introducing microwave flux bias lines.
A  lower bound for the time scale of the flux tuning is given by the plasma frequency of the SQUID, which is of the order of  30~GHz in our samples.
In the future,  qubits can  be integrated into this system enabling the demonstration the protocol for fast and accurate initialization~\cite{Tuorila_2017}.

\section*{Methods}

\noindent
\textbf{Theoretical model and simulations}\\
\noindent
We analyze the electrical circuit shown in Fig~\ref{fig:sample_structure}f, which also defines the symbols employed below.
The input impedance of Resonator 2 can be obtained from standard microwave circuit analysis~\cite{Pozar}, and it is given by
\begin{equation}
 \!  \!  Z_\textrm{r2} \! =  \! 
 \frac{1}{i \omega C_\textrm{T}} + 
 \frac{Z_0  \!  \left\{  \!  Z_\textrm{S}  \! + \!  Z_0  \!  \tanh(  \! \gamma x_2  \! )  \!  +   \frac{Z_0 \left[ Z_\textrm{term}  \!  +  \!  Z_0 \tanh(\gamma x_2) \right]}{Z_0  + Z_\textrm{term} \tanh(\gamma x_2)} \!  \right\}
}{
Z_0  \! +  \!  \tanh(\gamma x_2)  \left\{  \!  Z_\textrm{S}  \!  +  \!  \frac{Z_0 [ Z_\textrm{term} +  Z_0 \tanh(\gamma x_2) ]}{Z_0  +  Z_\textrm{term} \tanh(\gamma x_2)} \!   \right\}
},
\end{equation}
where $Z_\textrm{S}=i\omega L + 2/(i\omega C_\textrm{L})$ is the impedance of the SQUID and the parallel plate capacitors connecting the SQUID to the center conductor, 
and $Z_\textrm{term}=R+1/(i\omega C_\textrm{R1})+1/(i\omega C_\textrm{R2})$ is the impedance of the terminating resistor and the capacitances connecting it to the center conductor and the ground plane.
Here, $\omega=2\pi f$ is the angular frequency of the measurement tone, and $\gamma$ is the wave propagation coefficient detailed below.
We consider the SQUID as a tunable classical inductor. 
The  inductance of the SQUID as a function of the magnetic flux $\Phi$ is ideally given by
$ L(\Phi) = \Phi_0 /[2 \pi I_0 |\cos(\pi \Phi / \Phi_0)|],$
where $I_0$ is the maximum supercurrent through the SQUID, and $\Phi_0 = h/(2e)$ is  the magnetic flux quantum.
The losses in the SQUID are assumed to be substantially smaller than those induced by the resistor; thus, they are neglected.
One could also include a capacitance in parallel with the inductance in the model but it would have only a minor effect as discussed below.

We can calculate the scattering parameter from Port~1 to Port~2  using the ABCD matrix method~\cite{Pozar}
\begin{equation}
S_{21}=\frac{2}{ A  + B/Z_L  + C Z_L +  D},
\end{equation}
where the coefficients are obtained from
\begin{widetext}
\begin{equation}
 \begin{pmatrix}
 A   &B  \\
 C   &D
 \end{pmatrix} 
=
 \begin{pmatrix}
 1 &\frac{1}{i \omega C_\textrm{C}}\\
0 &1
 \end{pmatrix} 
 \begin{pmatrix}
 \cosh(\gamma x_1) & Z_0 \sinh(\gamma x_1) \\
\frac{1}{Z_0} \sinh(\gamma x_1)  & \cosh(\gamma x_1) 
 \end{pmatrix} 
 \begin{pmatrix}
 1 &0\\
\frac{1}{Z_\textrm{r2}} &1
 \end{pmatrix} 
 \begin{pmatrix}
 \cosh(\gamma x_1) & Z_0 \sinh(\gamma x_1) \\
\frac{1}{Z_0} \sinh(\gamma x_1)  & \cosh(\gamma x_1) 
 \end{pmatrix} 
 \begin{pmatrix}
 1 &\frac{1}{i \omega C_\textrm{C}}\\
0 &1
 \end{pmatrix}.
\end{equation}
\end{widetext}
These equations are solved numerically with Matlab.

The simulation parameters are given in Table~\ref{tab:simulation_parameters}.
We use  identical parameters in the simulations  for both samples except that the length $x_2$ is different.
The capacitances $C_\textrm{C}$ and $C_\textrm{T}$ are based on finite-element-method (FEM) calculations with design geometry and without native oxides, whereas $C_\textrm{L}$, $C_\textrm{R1}$, and $C_\textrm{R2}$ are calculated using parallel-plate-capacitor model by deducing the areas from scanning electron microscope  images, and assuming the niobium oxide  to have a typical thickness~\cite{Bach_thesis} of 5~nm and relative permittivity~\cite{Graca_2015} of 6.5.
The capacitance per unit length of the coplanar waveguide   $C_\textrm{l}$ is also based on a FEM simulation.
The resistance $R$ is measured with a dc control sample in a four-probe setup at 10~mK. 
The test resistor is fabricated in the same process with the actual samples.
The effective permittivity  of the  waveguide $\epsilon_\textrm{eff}$ is obtained from the nominal widths of the centre conductor and the gap,   10~$\mu$m and 5~$\mu$m respectively, using an analytical formula~\cite{Gevorgian_1995}.
The  lengths $x_1$ and $x_2$ are design values. 
The internal quality factor of the first mode of  Resonator 1 alone,  $Q_\textrm{int,1}$, is based on  measurements of  control samples consisting of a single resonator, and it agrees well with the measured  first mode of Sample~A.
The characteristic impedance of the external lines $Z_\textrm{L}$ has a nominal value, and the characteristic impedance of the resonators $Z_0$  has a design value in good agreement with the experimental results.
The maximum supercurrent through the SQUID $I_0$ is used as the only fitting parameter since it cannot be directly measured in the actual sample.
Nevertheless, the critical current in the actual samples is relatively close to a switching current of approximately 180~nA measured with a dc setup in an essentially similar but separately fabricated control SQUID. 
Due to noise from a high-temperature environment via the dc lines, the  temperature of  the control SQUID may be higher than in the actual sample, thus providing an explanation to the difference in the critical current and the switching current.
In addition, the difference  may well be explained by  unintentional differences in the fabrication.
We can write the wave propagation coefficient as
$ \gamma =  \omega_1 /(2 Q_\textrm{int,1} v_\textrm{ph}) + i\omega / v_\textrm{ph},$
where $v_\textrm{ph} = c/\sqrt{\varepsilon_\textrm{eff}}$ is the phase velocity, $\omega_1 /(2 \pi)=  c/(4 x_1 \sqrt{\varepsilon_\textrm{eff}})$ is the fundamental frequency,  and $c$ is the speed of light in vacuum.

The loaded quality factor can be defined as $Q_\textrm{L} = \omega_0 E / P_\textrm{loss}$, where $\omega_0 = 2 \pi f_0$ is the angular frequency of the resonance, $E$  the energy stored in the resonator, and $P_\textrm{loss}=-\textrm{d}E/\textrm{d}t$  the power loss.
Without input power, the energy in the resonator evolves as a function of time $t$ as  $E(t)=E_0 \exp(-\omega_0 t /Q_\textrm{L})$, where $E_0$ is the initial energy. 
Thus, the photon lifetime is given by $\tau_\textrm{L} = Q_\textrm{L}/\omega_0$, which corresponds to the total losses described by $Q_\textrm{L}$.
Since the number of photons in a resonator $n$ depends on the energy as $E=n \hbar \omega_0$, where $\hbar$ is the reduced Planck constant, and the power loss is bounded from above by the input power $P_\textrm{in}$ in the steady state, one obtains an upper bound  for the photon number as $n < Q_\textrm{L} P_\textrm{in} /(\omega_0^2 \hbar)$.
Therefore, the average photon number in a 10-GHz resonator is near unity or below  if the $Q$ factor is $10^5$ and the input power is $-140$~dBm (c.f. Fig.~\ref{fig:Qfactor_i7}).
The external quality factor corresponding to the leakage through the coupling capacitors can be calculated as\cite{Goppl} $Q_\textrm{ext} = 2 x_1 C_\textrm{l}  /(4   Z_\textrm{L} \omega_0 C_\textrm{C}^2)$. Although $Q_\textrm{ext}$ calculated with this formula is quite sensitive  to errors especially in $C_\textrm{C}$, it can be considered at least as an order-of-magnitude estimate.
The external quality factor is related  to  the coupling strength describing the external ports, $\kappa_\textrm{ext}=\omega_0/Q_\textrm{ext}= 2\pi \times 30$~kHz at $\omega_0 =  2\pi \times 10$~GHz.
In addition, one can write the photon lifetime without other loss mechanisms as $\tau_\textrm{ext} =  Q_\textrm{ext}/\omega_0= 1 / \kappa_\textrm{ext} = 6$~$\mu$s. 
The coupling to the external ports can be compared with the coupling strength  between the resonators at resonance  calculated as~\cite{Jones_4} $g_\textrm{T} =  C_\textrm{T} V_1 V_2 / \hbar =  2\pi \times 10$~MHz, where $V_{i} = \sqrt{\hbar \omega_0/(2 x_{i} C_\textrm{l})}$, $i=1,2$, and  $\omega_0 =  2\pi \times 10$~GHz.
Furthermore, the period for coherent oscillations between the resonators can be written as\cite{Blais} $\tau_\textrm{T} =  \pi / g_\textrm{T} = 30$~ns, where we have neglected dissipation.

The junction capacitance can be estimated using a parallel-plate model with an approximate aluminium oxide thickness of 2~nm, a junction area of 0.25~$\mu$m estimated from micrographs, and a typical relative permittivity~\cite{Landry_2001} of 8.2, which yield 10~fF per junction. 
At zero flux and 5~GHz (10~GHz), the inductive reactance of the SQUID is 40~$\Omega$ (80~$\Omega$) whereas the capacitor consisting of two junctions in parallel has a reactance of 2~k$\Omega$ (0.9~k$\Omega$).
If included in the model, the capacitive shunt of the inductance could result in  a very small change of the scattering parameter $S_{21}$ at $\Phi/\Phi_0\approx0.5$ where the inductance ideally diverges.
The change is small owing to the weak coupling of the resonators.
Consequently, we do not include it in the model.
Thus, the effect of the capacitance is effectively included in that of the inductance, which depends on the fitting parameter $I_0$.
The  plasma frequency of the SQUID can be obtained as $\omega_\textrm{p}/(2\pi) = 1/(2\pi\sqrt{LC})$, where $L$ is the inductance and $C$ the capacitance of the junctions.

\noindent
\textbf{Sample fabrication}\\
\noindent
Samples~A and B are fabricated  in the same process.
The actual samples as well as the control samples are fabricated on 100-mm Si wafers.
First, native SiO$_2$ is removed with ion beam etching, and 200~nm of Nb is sputtered onto the waver without breaking the vacuum.

Second, the large patterns are defined using standard optical lithography.
The optical lithography begins with hexamethyldisilazane  priming, followed by spin coating the resist AZ5214E at 4000 rpm.
The resist is exposed using a mask aligner in a hard-contact mode, and the exposed resist is removed with the developer AZ351B.
In order to obtain a positive profile for the Nb edges, we apply a reflow bake at 140$^\circ$C before reactive ion etching.
Once the large patterns are ready, we pre-dice the wafer half way from the back side.

In the third step, the nanostructures are defined using electron beam lithography (EBL).
After thorough cleaning of the wafer with a plasma stripper, a  resist for EBL is spin-coated to the wafer. 
The EBL resist consists of two layers: poly(methyl methacrylate) with 4\% of anisole, and poly[(methyl methacrylate)-co-(methacrylic acid)] with 11\% of ethyl lactate.
We fabricate all the nanostructures in a single EBL write.
For the development, we use a 1:3 solution of methyl isobutyl ketone and isopropanol.
The metallization for the nanostructures is carried out with an electron beam evaporator in two steps.
First, the Cu resistor is evaporated followed by the evaporation of the SQUID.
We evaporate Cu only on the area in the vicinity of the resistor on the chip and keep the rest of the chip covered by a metal mask. 
Subsequently,  we cover the resistor and evaporate the Al structures.
The SQUID consists of two Al layers evaporated at two angles ($\pm15^\circ$).
The oxide layer for the Josephson junctions is obtained by oxidizing Al in situ in the evaporation chamber at 1~mbar of O$_2$ for 5~min.
The lift-off process is carried out in acetone followed by cleaning with isopropanol.
The Cu resistor has a width of 250~nm, thickness of 30~nm, and length of 90~$\mu$m.
The SQUID consists of two layers of Al with thicknesses of 40~nm each, and it has a loop area of approximately 50~$\mu$m$^2$.

\noindent
\textbf{Measurement setup}\\
\noindent
The measurement setup is shown in Supplementary Fig.~\ref{fig:meas_setup}.
The measurements are carried out in a dry dilution refrigerator with a base temperature of approximately 10~mK, and the scattering parameters are measured with a vector network analyzer (VNA).
We control the magnetic flux through the SQUID using an external coil attached to the sample holder, and the current through the coil is generated with  a source-measure unit (SMU). 
The sample is wire-bonded to a printed circuit board shielded by a sample holder that is fabricated out of Au-plated Cu.
The sample holder is placed inside a magnetic shield to mitigate magnetic-field noise.

\noindent
\textbf{Normalization of scattering parameters}\\
\noindent
All raw experimental scattering parameters are normalized.
First, the winding of the phase as a function of frequency  is cancelled for convenience  by multiplying $S_{21}$ with $\exp(i2\pi f \tau)$ where $\tau\approx 50$~ns.
Second, the circle in the complex plane  drawn by $S_{21}$ when the frequency is swept through the resonance  is shifted and rotated to its canonical position, where the circle intersects the origin and the maximum amplitude lies on the positive $x$ axis~\cite{Petersan_1998}.
Any uncertainty in this shift causes relatively large errors near origin; hence, we use linear scale for experimental data as it emphasizes the large amplitudes with smaller relative error.
Consequently, one can extract the $Q$ factor using the phase--frequency fitting method discussed in Ref.~\onlinecite{Petersan_1998}.
In addition to the experimental $Q$ factors, we use the same method for obtaining the $Q$ factor also from the simulations, except that the very low $Q$ factor of Resonator~2 is obtained from the width of the dip.
In order to exclude uncertainty related to the cable losses, we normalize $S_{21}$ by dividing it with  $\max_{f,\Phi} |S_{21}|$ separately for each mode. 
The magnetic flux is extracted from the periodicity of the of modes 2 and 4, and there can be an irrelevant offset of an integer number of flux quanta.
One flux quantum corresponds to an electric current of approximately 2~mA in the coil used.



\noindent
\textbf{Acknowledgements} 
We acknowledge the provision of facilities and technical support by Aalto University at OtaNano - Micronova Nanofabrication Centre. 
We thank H. Grabert, M. Silveri, A. Wallraff, J. Kelly, J. Goetz and D. Hazra for discussions, and R. Kokkoniemi and S. Patom\"aki for technical assistance.
We have received funding from the European Research Council under Starting Independent Researcher Grant No. 278117 (SINGLEOUT) 
and under Consolidator Grant No. 681311 (QUESS), 
the Academy of Finland through its Centres of Excellence Program (project nos 251748 and 284621) and grants (Nos. 265675, 286215, 276528, 305237,  305306, 308161 and 314302), 
 the Vilho, Yrj\"o and Kalle V\"ais\"al\"a Foundation,
the Technology Industries of Finland Centennial Foundation,
and the Jane and Aatos Erkko Foundation.

\noindent
\textbf{Author Contributions}
M.P. was  responsible for sample design, fabrication, measurements, data analysis, and for the writing of the initial version of the manuscript.
K.Y.T and J.G. contributed to the sample design, fabrication, measurements, and analysis. 
S.M. contributed to the analysis, and M.J. to the measurements and analysis.
L.G. deposited the Nb layer.
M.P., K.Y.T, J.G, R.E.L., L.G., J.H., S.S., V.V., and M.M. developed the fabrication process.
J.T. and T.A.-N. contributed to the theoretical understanding of the system.
M.M. provided initial ideas and supervised the project.
All authors commented on the manuscript.

\noindent
\textbf{Competing financial interests}  The authors declare no competing financial interests.

\noindent
\textbf{Data availability} The data is available upon  request from the  authors.

\noindent
\textbf{Supplementary material}
Supplementary Figures \ref{fig:meas_sim_modes_i8}, \ref{fig:meas_sim_delta_f}, \ref{fig:Qfactor_i8}, \ref{fig:meas_setup}.

\renewcommand{\figurename}{Supplementary Figure}
\renewcommand{\tablename}{Supplementary Table}

\setcounter{figure}{0} 
\setcounter{table}{0} 

\begin{figure*}[hb]
\centering
\includegraphics[width=\linewidth]{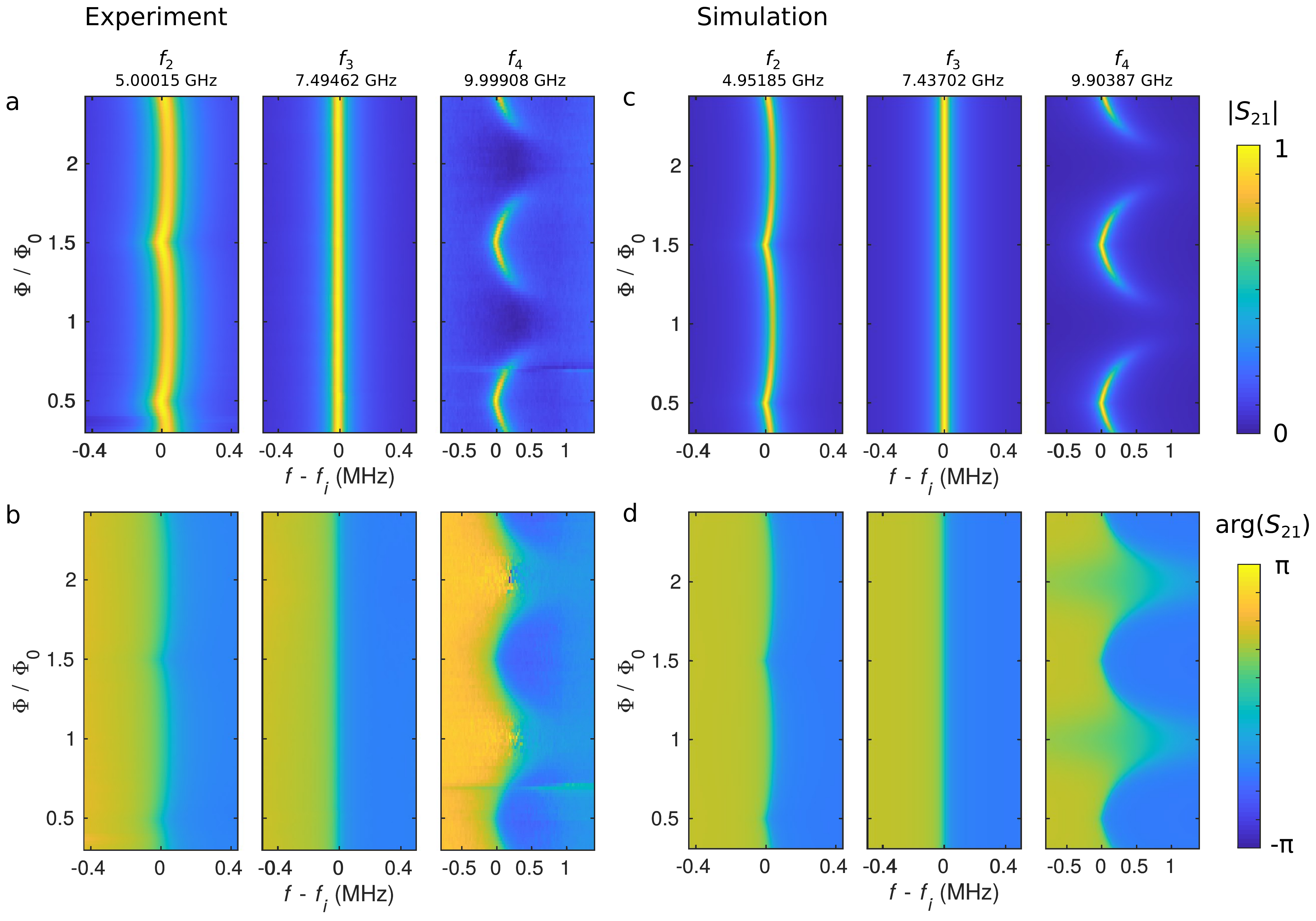}
\caption{Resonances of  Sample~B. (a,b) Experimental and (c,d) computational scattering parameter $S_{21}$ for the  modes 2,3, and 4 of Resonator 1  as  functions of frequency and magnetic flux.
(a,c) Normalized  amplitude of $S_{21}$. Each panel is normalized separately by dividing with the corresponding maximum amplitude.
(b,d) Phase of $S_{21}$. 
The resonance frequencies are given above the panels, and the simulation parameters are given in Table~\ref{tab:simulation_parameters}.
The power in the experiments is approximately $-90$~dBm at Port~1.
}
\label{fig:meas_sim_modes_i8}
\end{figure*}

\begin{figure*}[t]
\centering
\includegraphics[width=0.85\linewidth]{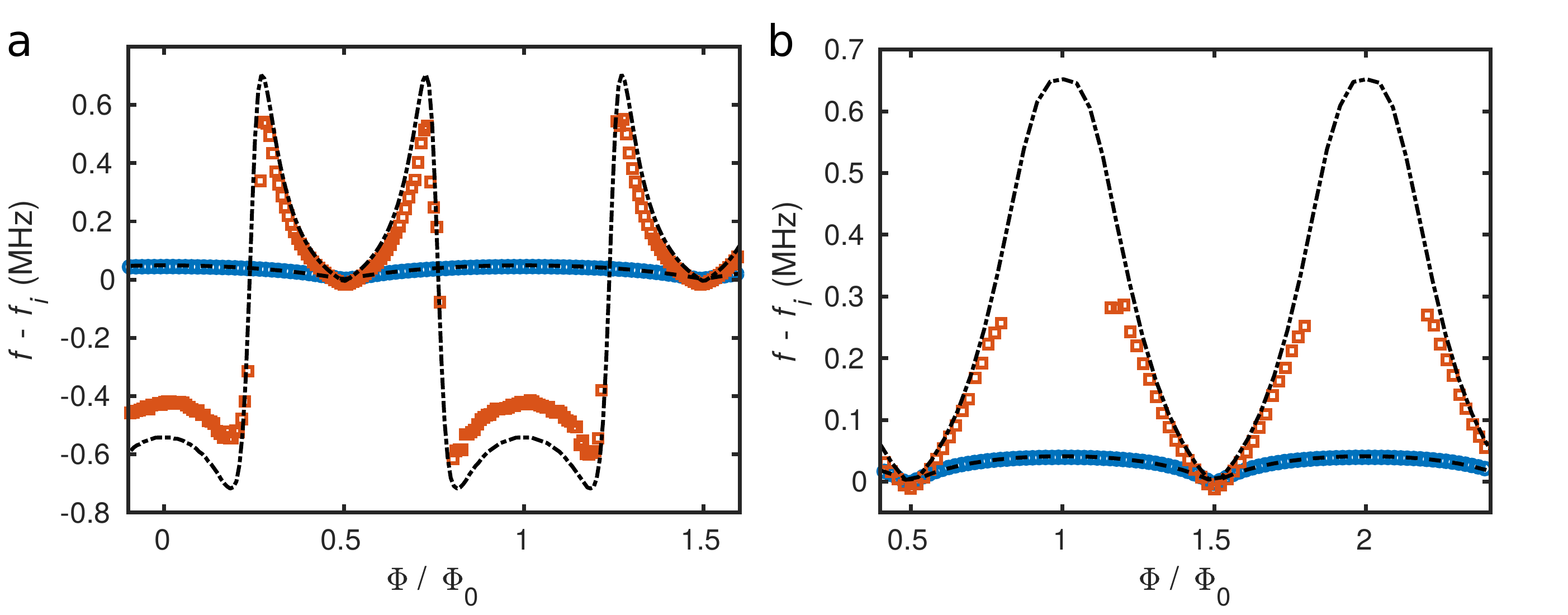}
\caption{
Resonance frequency shifts from the magnetic flux point $\Phi/\Phi_0=0.5$. 
(a) Measured frequency shifts of modes 2 (blue circles) and 4 (red squares) together with the corresponding simulations for modes 2 (dashed line) and 4 (dash-dotted line) of Sample~A as functions of the magnetic flux.
(b) As (a) but for Sample~B.
The simulation parameters are given in Table~\ref{tab:simulation_parameters}.
For the frequencies $f_{2/4}$ in Samples~A and B, see Fig.~\ref{fig:meas_sim_modes_i7} and Supplementary Fig.~\ref{fig:meas_sim_modes_i8}, respectively.
}
\label{fig:meas_sim_delta_f}
\end{figure*}

\begin{figure*}[t]
\centering
\includegraphics[width=0.85\linewidth]{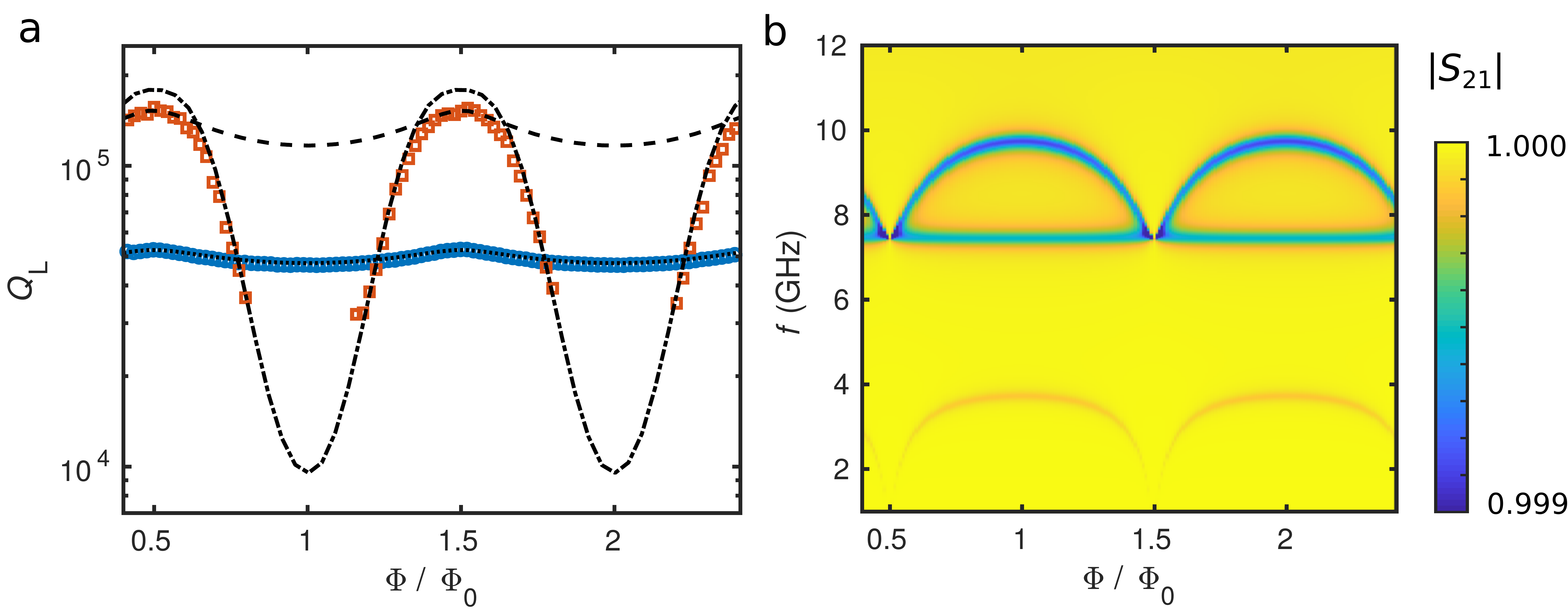}
\caption{Quality factors of Resonator 1 and resonances of Resonator 2 for Sample~B.
(a) Measured loaded quality factor, $Q_\textrm{L}$, for mode~2 (blue circles), and for mode 4 (red squares) as functions of the magnetic flux through the SQUID together with the simulated values (dashed line and dash-dotted line, respectively). 
The dotted line on top of the blue circles shows the simulation for the mode 2 with an additional spurious loss mechanism with a flux-independent quality factor  of $Q_\textrm{sp}=8\times10^4$ in addition to the simulated  quality factor, $Q_\textrm{L,si}$,  yielding  $Q_\textrm{L}^{-1} = Q_\textrm{L,si}^{-1} + Q_\textrm{sp}^{-1}$ with a better match with the experimental data.
The applied power is approximately $-90$~dBm at Port~1.
(b) Absolute value of the simulated scattering parameter $S_\textrm{21}$ with only Resonator~2, i.e., at the limit $C_\textrm{C}\rightarrow \infty$. 
The colour bar is truncated at 0.999 for clarity. 
The simulation parameters are given in Table~\ref{tab:simulation_parameters}.
}
\label{fig:Qfactor_i8}
\end{figure*}

\begin{figure*}[p]
\centering
\includegraphics[width=0.5\linewidth]{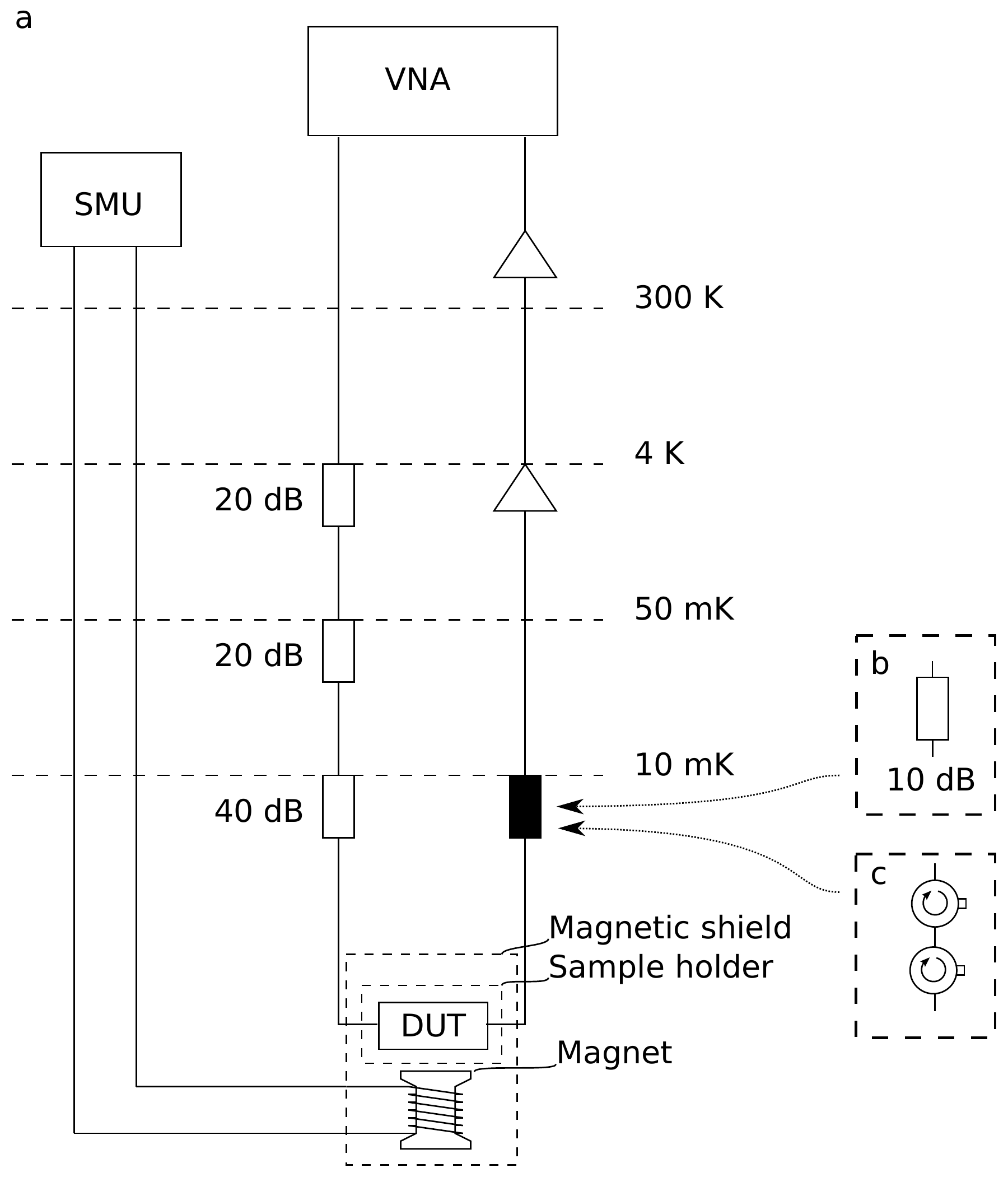}
\caption{Measurement setup.
(a) Overview of the measurement setup with different temperature stages of the cryostat indicated. 
The scattering parameters of the device under test (DUT) are measured with a vector network analyzer (VNA), and the magnetic flux through the SQUID is produced with a bias current generated by  a source measure unit (SMU). 
(b) For Sample A, a 10-dB attenuator is employed after the sample in the position of the black box to prevent amplifier noise from entering the sample. 
(c) For Sample B, two isolators are used instead.
}
\label{fig:meas_setup}
\end{figure*}

\clearpage

\end{document}